\documentclass[12pt,english]{iopart} 
\usepackage[T1]{fontenc}
\usepackage[latin9]{inputenc}
\usepackage{float}
\usepackage{amsthm}
\usepackage{bbold}
\usepackage{mathdots}
\usepackage{stackrel}
\usepackage{graphicx}
\usepackage{color}
\usepackage{titlesec}

\titleformat{\section}[runin]{\bfseries\itshape}{}{0pt}{}[.---]
\titleformat{\subsection}[runin]{\itshape}{}{0pt}{}[.---]
\makeatletter

\newtheorem{prop}{Proposition}

\newcommand{\kon}[1]{#1}
\newcommand{\ko}[1]{#1}

\newcommand{\text}[1]{\mathrm{#1}}
\newcommand{\eqref}[1]{(\ref{#1})}
\newcommand{\eva}[1]{\ensuremath\left\langle {#1} \right\rangle}
\makeatother

\usepackage{babel}
\begin{document}


\title{Geometric and algebraic origins of additive uncertainty relations}

\author{Konrad Szyma\'{n}ski$^1$, Karol \.{Z}yczkowski$^{1,2}$ }

\date{May 14, 2018}

\address{$^1$ Marian Smoluchowski Institute of Physics, Uniwersytet Jagiello\'{n}ski, Krakow, Poland}

\address{$^2$ Center for Theoretical Physics of the Polish Academy of Sciences, Warszawa, Poland}

\begin{abstract}
Constructive techniques to establish state-independent uncertainty relations for the sum of variances of arbitrary two observables are presented.
We investigate the range of simultaneously attainable pairs of variances,
which can be applied to a wide variety of problems including finding exact bound for the sum of variances of two components of angular momentum operator for any total angular momentum quantum number $j$ and detection of quantum entanglement.
Resulting uncertainty relations are state-independent, semianalytical, bounded-error 
and can be made arbitrarily tight. 
The advocated approach, based on the notion of joint numerical range of a number of observables and uncertainty range, allows us to improve earlier numerical works and to derive semianalytical tight bounds for the uncertainty relation for the sum of variances  expressed as roots of a polynomial of a single real variable.
\end{abstract}
\maketitle

\section{Introduction}
Uncertainty relations form a
 wide branch of problems with {ubiquitous} applications: from the {tests} of quantum theory \cite{scully1991quantum} through quantum cryptography \cite{tomamichel2012tight} and entanglement detection \cite{hofmann2003violation,guhne2004characterizing} to direct usage in experiments \cite{sorensen2001entanglement}. There exist various 
approaches to the problem: the original articles by {Heisenberg and Kennard \cite{kennard1928note,heisenberg1930physikalische}}
 consider the product of variances of two observables, 
while a recent approach by Maccone and Pati \cite{maccone2014stronger} 
\kon{concerns} the 
sum of variances, especially useful for finite-dimensional systems. Entropic uncertainty relation \cite{bialynicki1984entropic} {bounds} from below 
the sum of entropies of probability distributions of observed quantities.
In the state-independent approach for a selected  
pair of observables one derives \kon{lower} bounds for a given quantity
valid for any quantum state.
These results can be often improved in the state-dependent approach,
in which one establishes more precise dedicated bounds, which
depend explicitly on the measured quantum state.

Current interest in improving uncertainty relations 
\cite{maccone2014stronger,schwonnek2017state,giorda2018state,sehrawat2017deriving,dammeier2015uncertainty} 
is motivated by their numerous applications in mathematical physics
and \kon{in} the theory of quantum information processing.
The goal of this work is to provide a novel, 
geometrical view on the variance-based uncertainty relations, which
allows one to establish exact analytical results.
The current contribution extends recent works
\cite{schwonnek2017state,szymanski2017uncertainty},
in which approximate uncertainty relations for the sum of variances
were obtained with help of numerical techniques.

We provide here semianalytical, bounded-error,
state-independent bounds for uncertainty relations involving variances of two operators and analyze properties of an associated geometric object, called {\sl uncertainty range}.
Furthermore, we propose a procedure determining 
a tight, state-independent bound for the sum of variances as a root of a certain polynomial. While the roots of \kon{the algorithmically generated} polynomials always exist, usually the procedure is feasible in low dimensions and leads to exact, explicit analytic bounds. The procedure is exemplified by analytical determination of minimal sum of variances for angular momentum operators, $\Delta^2 J_X + \Delta^2 J_Y$ for several values of the total angular momentum quantum number $j$, which expands on known numerical results \cite{dammeier2015uncertainty}.

\section{Sum-of-variances uncertainty relations} 

To characterize uncertainty related to a double quantum measurement
one can analyze the sum of variances \cite{maccone2014stronger}.
For any two operators $X$ and $Y$ we wish to provide a 
state-independent bound $C(X,Y)$ for the sum of variances:
\begin{equation}
\Delta^{2}X+\Delta^{2}Y\ge  C(X,Y).
\label{eq:firstsum}
\end{equation}

One of the possible approaches makes use of the following fact:
it is possible to rewrite the sum of variances as a function of the averages,
\begin{equation}
\Delta^{2}X+\Delta^{2}Y=\langle X^{2}+Y^{2}\rangle-\langle X\rangle^{2}-\langle Y\rangle^{2},
\label{eqn:ump}
\end{equation}
so it is sufficient to minimize the function $g=\left\langle X^2+Y^2\right\rangle 
 -\left\langle X\right\rangle ^{2}-\left\langle Y\right\rangle ^{2}$ over all states. If one could determine the set of allowed triples of expectation values $(\langle X\rangle,\langle Y\rangle,\langle X^2+Y^2\rangle)_{\rho}$, the function $g$ could be interpreted as a simple polynomial of three variables, which is minimized over set of triples. This can be done, and numerical methods employing this observation have been recently developed \cite{schwonnek2017state,szymanski2017uncertainty}, here we present an analytical extension allowing for strict treatment of several interesting classes of observables.

\section{Numerical range of observables}
In {this work on uncertainty relations we will use the notion of \emph{numerical range} -- the set of simultaneously allowed expectation values}. In this section the most important properties necessary for discussion of uncertainty relations are presented.

Expectation value of an observable $F$ on a pure state,
$\langle F\rangle_{\psi}=\langle \psi |F|\psi\rangle$,
is a key concept of the quantum theory.
For any hermitian matrix $F$ of a fixed order $d$
one can pose the question,
what is the range of possible expectation values
among all normalized pure states: the answer is a segment {of the real axis bounded by the extremal eigenvalues of $F$}, $[\lambda_{\text{min}},\lambda_{\text{max}}]$.

A similar problem was earlier analyzed
by mathematicians, who studied an algebraic notion 
of  {\sl numerical range} of an (not necessarily hermitian) operator $X$ --
a subset $W$ of the complex plane defined by 
\begin{equation}
W(X)=\{ z{\in\mathbb{C}}:\  z= \langle \psi |X|\psi\rangle, \
\langle \psi |\psi\rangle =1 \}.
\label{WX}
\end{equation}
\ko{Numerical range $W(X)$ may be interpreted as a set of allowed expectation values of a single operator $X$.}
A classical theorem of Toeplitz and Hausdorff,
obtained nearly a century ago 
\cite{toeplitz1918algebraische,hausdorff1919wertvorrat},
states that for any matrix $X$ the set $W(X)$ is convex.
If {the} matrix $X$ is normal then the set $W(X)$ forms the convex hull of the spectrum of $X$.
For a hermitian observable {$F$ the corresponding set $W(F)$} reduces to an interval belonging to the
real axis \cite{horn1990matrix}.
For operators of dimension $d\le 3$ 
the possible shapes of numerical ranges are 
classified \cite{keeler1997numerical,helton2011possible}.

Expectation value of \emph{single} operator may not capture the whole complexity of some problems. Furthermore, in the definition in Eq. \eqref{WX}, only pure states are taken into account, while the set of mixed states is potentially much more intricate. A natural generalization to \emph{sequence of $k$ averages} of fixed operators $(F_i)_{i=1}^k$ taken over \emph{all mixed states} is used instead. The set of $k$ expectation values which can be obtained by measuring $k$ Hermitian observables {$(F_i)_{i=1\ldots k}$} over a common quantum state $\rho$ is called \emph{joint numerical range} \cite{li2000convexity} and defined by
\begin{equation}
\eqalign{
W(F_1,F_2,\dots , F_k)=\{ {\bf x}&\in {\mathbb R}^k : x_j=
\langle F_j\rangle_{\rho}, 
\; j=1,\dots, k; \; \rho\in {\cal M}_d \},}
\label{WFF}
\end{equation}
where 
$\langle F\rangle_{\rho} ={\rm Tr} \rho F$. {Here }
${\cal M}_d =\{\rho: \; \rho^{\dagger}=\rho\ge 0; \; {\rm Tr}\rho=1\}$
denotes the set of all normalized mixed states of size $d$. The additional motivation for taking mixed states into account is to ensure convexity of the resulting set: evidently, the set of expectation values of three Pauli matrices \emph{among pure states} forms the hollow Bloch sphere.
As any operator can be decomposed into its hermitian and antihermitian part,
$X=F_H+i F_A$, the set $W(X)$ can be considered
as a {\sl joint numerical range} (JNR) of two hermitian observables, $W(F_H,F_A)$.
This is the set of simultaneously attainable expectation values of these 
two observables obtained in a double quantum measurement
performed on two copies of the same state.

A classification of 3D numerical ranges of a triple
of operators of size $d=3$ was given in \cite{szymanski2016classification}.
JNR and its extensions {find diverse applications, }the notion
allows for visualization of phase transitions \cite{chen2016physical,zauner2016njp,spitkovsky2017new},
{construction} of nonlinear entanglement witnesses or to improve our
understanding of the geometry of the set of of quantum states, as joint numerical range describes its projection onto lower dimensional subspaces (up to affine transformations). 
 \cite{szymanski2016classification}.

\section{Reformulated definition of variance}
The methods used in this paper strongly rely on an equivalent definition of variance, which allows for efficient analytical treatment and -- if necessary -- {is easy to approximate}. This is achieved by defining the variance over {the} specific state $\Delta^2 X_\rho$ by {minimal} expectation value taken over the set of linear functions of $\langle X^2\rangle$ and $\langle X\rangle$. This equivalent definition of variance {reads}
\begin{equation}
\Delta^2 X_\rho := \min_{x\in\mathbb{R}} \langle (X-x\mathbb{1})^2\rangle_\rho = \min_{x\in\mathbb{R}}~\langle X^2 - 2x X + x^2 \mathbb{1}\rangle_\rho,
\end{equation} 
that is, {the} minimum taken over {the}s \emph{set of linear functions} of operators $X^2$ and $X$ parametrized by $x\in\mathbb{R}$. In order to calculate, for instance, {a} lower bound {for the} sum of variances, both {terms can be} simultaneously optimized,
\begin{equation}
\min_\rho \Delta^2 X_\rho+\Delta^2 Y_\rho = \min_\rho \min_{x,y\in\mathbb{R}}~\langle X^2+Y^2 - 2(x X+yY) + (x^2+y^2)\mathbb{1}\rangle_\rho.
\label{eq:form2}
\end{equation} 
Analytical treatment of this formulation is demonstrated in the next section.

Efficient approximations are made possible by selecting other (finite) set $S$ of operators which are linear in $X$ and $X^2$. 
{This method is discussed in details} after the next section.
\section{Application to uncertainty relations}
Exchanging the order of minimisation in Eq. \eqref{eq:form2} does not change the value of this expression, as we are looking for global minima. Then the expression reads,
\begin{equation}
\min_\rho \Delta^2 X_\rho+\Delta^2 Y_\rho =  \min_{x,y\in\mathbb{R}}\min_\rho~\langle X^2+Y^2 - 2(x X+yY) + x^2+y^2\rangle_\rho,
\label{eq:form3}
\end{equation} 
with the notation that the operator proportional to identity $(x^2 +y^2) \mathbb{1}$ is denoted by the proportionality factor only, ($x^2+y^2$); this notation is going to be used in the following text.

For a fixed operator $Z$ it is known {that the minimal value of the expectation value $\min_\rho ~\langle Z\rangle_\rho$ is equal to the minimal eigenvalue of $Z$}. Let us denote it by $\lambda_{\min} (Z)$ and write
\begin{equation}
\min_\rho \Delta^2 X_\rho+\Delta^2 Y_\rho =  \min_{x,y\in\mathbb{R}}~\lambda_{\min}\left( X^2+Y^2 - 2(x X+yY)+x^2+y^2\right).
\label{eq:form4}
\end{equation} 
We are now looking for a minimal eigenvalue of a {certain family of operators. This is a problem well suited for treatment with the theory of solving a system of polynomial equations:   
the conditions for minima in $x,y$ in Eq. \eqref{eq:form4} can be written as derivatives of characteristic polynomial (i.e. the minima are smooth and not cusp-like, see Appendix A). As a result the set of three polynomial equations in three real variables: $\lambda,x,y$ is obtained,}
\begin{equation}
\left\{\begin{array}{rl}
D:=\det \left(X^2+Y^2 - 2(x X+yY)+(x^2+y^2-\lambda)\right) & =0,\\
\partial_x D&=0,\\
\partial_y D&=0.
\end{array}\right.
\label{eq:f2d}
\end{equation}
These three equations always have a common solution, furthermore: the polynomials defining the solutions can always be found by algorithms used to determine the {Gr\"obner basis}. This means that a polynomial $R(\lambda)$, which minimal real root (with corresponding $x,y$ being real as well) is the lower bound of the additive uncertainty relation,
\begin{equation}
\min_\rho \Delta^2 X_\rho+\Delta^2 Y_\rho =\min \left\{ c: R(c)=0\text{; corresponding }~x,y~\text{ are~real}\right\}.
\label{eq:form5}
\end{equation} 

Geometric interpretation of this method reveals a close relation to the 3-D joint numerical range $W(X,Y,X^2+Y^2)$. If a triple $(\langle X\rangle,\langle Y\rangle,\langle X^2+Y^2\rangle )$ belongs to the set $W(X,Y,X^2+Y^2)$, the sum of variances, {denoted by $v$, can be rewritten as} 
\begin{equation}
v(\langle X\rangle, \langle Y\rangle, \langle X^2+Y^2\rangle)=\langle X^2+Y^2\rangle-\langle X\rangle^2-\langle Y\rangle^2.
\end{equation}
Is it clear that the surfaces of equal variance are paraboloids of revolution. If we now introduce one of the functions $w_{x,y}$ we are minimizing over,
\begin{equation}
w_{x,y}(\langle X\rangle, \langle Y\rangle, \langle X^2+Y^2\rangle)=\langle X^2+Y^2\rangle - 2(x \langle X\rangle+y\langle Y\rangle) + x^2+y^2,
\end{equation}
it can be easily concluded that $w_{x,y}(\langle X\rangle, \langle Y\rangle, \langle X^2+Y^2\rangle)=(\langle X\rangle, \langle Y\rangle, \langle X^2+Y^2\rangle)$ is equivalent to $x=\langle X\rangle, y=\langle Y\rangle$ with no restrictions on $\langle X^2+Y^2\rangle$. Hence, the function $w_{x,y}$ calculates true sum of variances if restricted to the straight vertical line; elsewhere it is a strictly upper bound.

Using this property we conclude that {if a global minimum of $w_{x,y}$ is attained for certain $x=x_0, y=y_0$ with triple of arguments $(\langle X\rangle,\langle Y\rangle, \langle X^2+Y^2\rangle)$, then the triple corresponds to minimal global sum of variances.}

\begin{figure}[t]
\centering
\includegraphics[width=0.6\columnwidth]{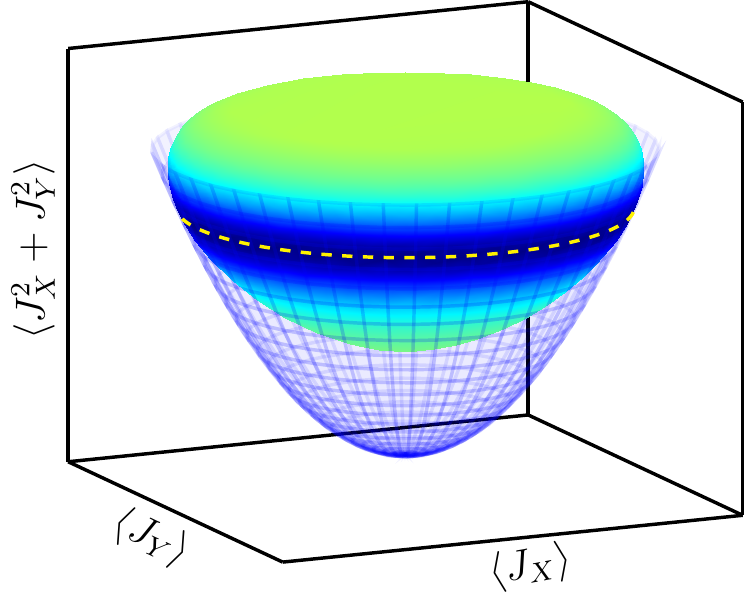}
\caption{3D numerical range $W(J_X,J_Y,J_X^{2}+J_Y^{2})$ for total angular momentum $j=3/2$ shaded according to sum of variances $\Delta^2 J_X+\Delta^2 J_Y$.
Paraboloid of a minimal uncertainty (sum of variances) is shown, as well as the minimizing manifold (yellow dashed circle). The linear approximator to the variance $w_{x,y}$ attains the same global minimum on the same set.}
\label{fig:uncermac}
\end{figure}
\subsection{Examples}
{1) \bf{Bounds for angular momentum operators.}}
The method described above allow us to obtain new analytical bounds for sum of variances of angular momentum. This particular problem enjoys additional symmetry, which greatly reduces the computational load: since $W(J_X,J_Y,J_X^2+J_Y^2)$ has rotational symmetry (see Fig. \ref{fig:uncermac}), we can set either of the variables $x,y$ in Eq. \eqref{eq:f2d} to $0$ and we are left with {the} set of 2 polynomial equations with 2 variables (see Appendix A for the  explicit form of polynomials for $j\le 4$). Making use of the techniques described above we obtained new, tight and analytical bounds for a pair of angular momentum operators $X=J_X$ and $Y=J_Y$ for arbitrary value of total angular momentum quantum number $j$. The data presented in Table \ref{tab:angmom} corresponding to tight and exact values of additive uncertainty relation bounds constitute a key result of this work.

\begin{table}
\begin{center}
\begin{tabular}{c|c|c |||c|c|c}
Total angular\\momentum $j$ & $\min \Delta^2 J_X + \Delta^2 J_Y$  & polynomial order &  $j$ & $\min \Delta^2 J_X + \Delta^2 J_Y$ & order \\\hline \hline \hline
${1}/{2}$&	$1/4=0.25$&	1&	$1$&	$7/16=0.4375$&	1\\\hline
${3}/{2}$&	$\approx0.6009$&	3&	$2$&	$\approx0.7496$&	3\\\hline
${5}/{2}$&	$\approx0.8877$&	7&	$3$&	$\approx1.018$&	    6\\\hline
${7}/{2}$&	$\approx1.142$&	    13&	$4$&	$\approx1.260$&	    10\\\hline
${9}/{2}$&	$\approx1.374$& 	21&	$5$&	$\approx1.484$&	    15\\\hline
${11}/{2}$&	$\approx1.591$&	    31&	$6$&	$\approx1.695$&	    21\\\hline
${13}/{2}$&	$\approx1.796$&	    43&	$7$&	$\approx1.894$&	    28\\\hline
${15}/{2}$&	$\approx1.991$&	    57&	$8$&	$\approx2.085$&	    36\\\hline
${17}/{2}$&	$\approx2.178$&	    73&	$9$&	$\approx2.268$&	    45\\\hline
${19}/{2}$&	$\approx2.358$&	    91&	$10$&	$\approx2.445$&	    55
\end{tabular}
\end{center}
\label{tab:angmom}
\caption{Tight and analytial bounds for state-independent uncertainty relation involving sum of variances of angular momentum operator for varying total angular momentum quantum number $j$. Until now, only numerical results were available for $j>1$ \cite{dammeier2015uncertainty}. Sequence of orders of minimal polynomials is recognized by OEIS \cite{oeis} as A243099 -- see Appendix D for details.}
\end{table}

B) {\bf{Family of uncertainty relations.}}
Consider a family of uncertainty relations derived from angular momentum operators for $j=1$:
\begin{equation}
\Delta^2 J_X + \alpha \Delta^2 J_Y \ge C(\alpha).
\end{equation}
The bound for weighted sum of variances can be determined in the same way as previously -- 
there is no conceptual change in the calculations apart from the fact that the number of variables is higher -- leading to a piecewise function,
\begin{equation}
C(\alpha)=\left\{\begin{array}{cc}
\frac12-\frac1{16\alpha}: & \alpha\ge 1\\
 \frac{\alpha}2-\frac{\alpha^2}{16}: & \alpha\le 1
\end{array}\right..
\end{equation}

C) {\bf Arbitrary operators.}
The method described above is not limited to angular momentum operators only. Using the same technique we can determine, for instance, the minimal sum of variances for operators
\begin{equation}
X=\left(\begin{array}{ccc}
-1&0&0\\
0&0&0\\
0&0&1
\end{array}\right)~~\text{and}~~Y=\left(\begin{array}{ccc}0&1&0\\1&0&i\\0&-i&0\end{array}\right).
\label{eq:nonmom}
\end{equation}
The polynomial arising from solving Eq. \eqref{eq:f2d} leads to strict bound for sum of variances $\Delta^2 X+\Delta^2 Y\ge C=15/32$. The 3-D numerical range corresponding to this pair of operators is presented in Fig. \ref{fig:uncermac2} in Appendix A.
\section{Uncertainty range}

In analogy to numerical range $W(X,Y)$ defined in \eqref{WFF}, the set of possible expectation values,
one can introduce a geometric object containing information 
about simultaneously possible \emph{variances} of two operators,
\begin{equation}
U(X,Y) : =\left\{ \left(\Delta_{\rho}^{2}X,\Delta_{\rho}^{2}Y\right)\in \mathbb{R}^2:\rho\in\mathcal{M}_d\right\},
\label{eqn:UR}
\end{equation}
which will be called \emph{uncertainty range} -- see \cite{dammeier2015uncertainty,schwonnek2017state} for examples of application.

Uncertainty range is a nonlinear transformation of {the} 4-dimensional numerical range $W(X,X^2,Y,Y^2)$ and {in general it is not} a convex set -- see Fig. \ref{fig:uraprox}. 
Note that uncertainty range contains information about all additive uncertainty relations $\Delta^2(a X)+\Delta^2(b Y)\ge C(a,b)$. The bound $C(a,b)$ is determined by the line with normal $(-b,a)$ tangent to the uncertainty range.

\subsection{Sector decomposition}
In this section we are going to present the procedure of approximate variance; it is able to generate a set of operators $S$ defining approximation to variance through the relation
\begin{equation}
\label{eq:ap2}
\Delta^2 X_\rho \approx \min_{V\in S} \langle V \rangle_\rho.
\end{equation}
It enjoys several favorable characterstics: the approximation is a lower bound; its error is bounded by above and easily controlled.

\begin{figure}[H]
\centering
\includegraphics[width=0.9\columnwidth]{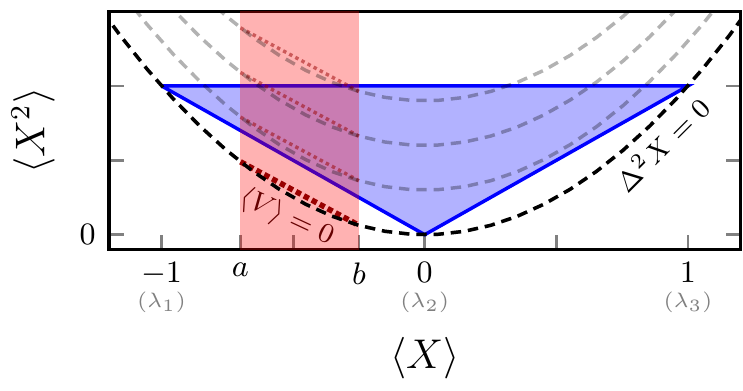}
\caption{Visualization of the sector decomposition for an operator $X$ of size 3 with spectrum $(\lambda_1,\lambda_2,\lambda_3)=(-1,0,1)$, for instance $J_X$ for total angular momentum quantum number $j=1$. Numerical range $W(X,X^2)$ (blue) is convex hull of points corresponding to eigenvalues of $X$ (corners of the triangle). Shaded in red is a \emph{sector} between $a$ and $b$ where a linear function of 
$\langle X^2\rangle$ and $\langle X\rangle$ (with dotted contour lines) approximates $\Delta^2 X$ (with dashed contour lines) from below. 
Entire set $W(X,X^2)$ can be covered by similar sectors.

}
\label{im:sec}
\end{figure}

We are going to use the spectral structure of the arbitrary analyzed observables $X$ and $Y$. 
Observe that the numerical range $W(X,X^{2})$ is determined by the spectrum 
of $X$: since operators $X$ and  $X^2$ do commute, the range $W(X,X^2)$ 
is a polygon formed by the convex hull of points $ (\lambda_i,\lambda_i^{2})$,
where $\lambda_i$ denote eigenvalues of $X$ 
-- see Fig. \ref{im:sec}.
 The aim is to provide a linear approximation 
${f}(\langle X\rangle,\langle X^2\rangle)=\alpha \eva{X}+\beta\eva{X^2}+\gamma$
to the true variance $v(\eva{X},\eva{X^2})=\eva{X^2}-\eva{X}^2$\emph{ valid} 
in some subset of $W(X,X^{2})$ in
the sense: $f$ approximates $v$ nontrivially from below, 
$0\le {f}(\eva{X},\eva{X^2})\le {v}(\eva{X},\eva{X^2})$.

No single choice of linear function is \emph{valid} in the entire set
$W(X,X^{2})$. To obtain meaningful results this set has to be covered
with validity regions of finite numbers of approximations. Natural choice is to split the whole region into vertical bands $a\le \langle X\rangle \le b$, such that the approximation is exact on $\langle X\rangle=a$ and $\langle X\rangle=b$. By minimizing \kon{the maximal} possible error, $u-f$, we are left with equation $f(\eva{X},\eva{X^2})=-(a+b)\eva{X}+\eva{X^2}+ab$.

\begin{figure}[H]
\centering
\includegraphics[width=0.6\columnwidth]{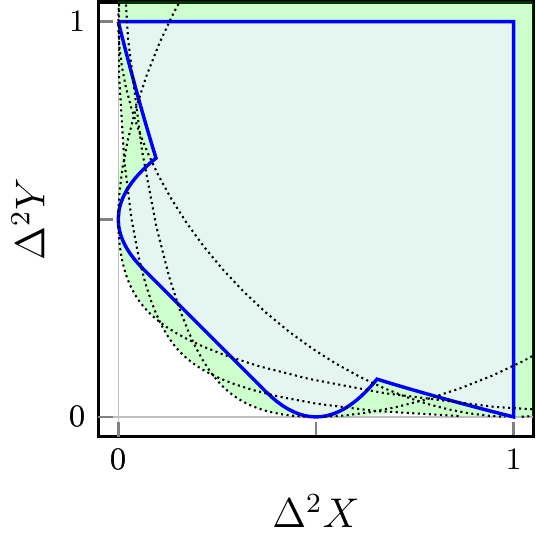}

\caption{Exemplary uncertainty range (light blue with blue boundary) $V(X,Y)$ and partial approximations generated by joint numerical ranges $W(X_i,Y_i)$ (with boundary depicted by dotted lines) and the total approximations (light green) --
 union of partial approximations. The nonconvexity of uncertainty range and approximation is visible.

In this figure operators $X$ and $Y$ are angular momentum operators for $j=1$: $X=J_X$ and $Y=J_Y$. }

\label{fig:uraprox}
\end{figure}
The minimal choice of the bands 
corresponds to an approximation of the variance with $d-1$ linear
functions between adjacent eigenvalues: for ordering $\lambda_1\le\lambda_2\le\ldots\le\lambda_d$, parameters of $i$-th function are: $a_i=\lambda_{i}$ and $\quad b_i=\lambda_{i+1}$. 
Denoting the maximum spacing between adjacent eigenvalues by $s_{\text{max}}=\max_{i}(\lambda_{i}-\lambda_{i+1})$, 
we find that the maximum error of variance approximation reads
$\delta_X=(s_{\text{max}})^2/4$.

Let us denote the sector decomposition of two operators $X,$ $Y$ by
$\left\{ X_{i}\right\} $, $\left\{ Y_{i}\right\}$. Then we can
provide ``a lower approximation'' to the uncertainty range $V(X,Y)$
by
\[
\bar{V}=\bigcup_{i,j}W(X_{i},Y_{j}), \ \ i,j=1,\dots, d-1,  
\]
where the ``lower approximation'' is to be understood in the following
way: if at each point $(x,y)\in\bar{V}$ we attach a rectangle $E:=[x,x+\delta_X]\times[y,y+\delta_Y]$ ($\delta$ being the maximum error of the variance approximation),
the uncertainty range is contained in the resulting set, Minkowski sum \cite{hadwiger1950minkowskische} of $\bar V$ and $E$ (see Fig. \ref{fig:uraprox}):

\[
V(X,Y)\subset\bar V \oplus E = \left\{ v+e : v\in \bar V, e\in E\right\}
\]

This construction is compatible with a nonconvex structure of the uncertainty range $V(X,Y)$ -- see Fig. \ref{fig:uraprox}. Such a procedure provides simple bounds for
the sum-of-variances uncertainty relation: a state-independent bound
reads

\[
\Delta^{2}X+\Delta^{2}Y\ge\min_{i,j}\left[\lambda_{\text{min}}(X_{i}+Y_{j})\right].
\]

Additionally, the error of approximation is at most the sum of sector decomposition errors of $X$
and $Y$, so in case of the minimal selection the difference of approximation and real bound is

\begin{equation}
\eqalign{
\Big|\big(\min\Delta^2X +\Delta^2Y&\big)-\min_{i,j}\big[\lambda_{\text{min}}(X_{i}+Y_{j})\big]\Big|\le \delta_X +\delta_Y.}
\end{equation}

The above reasoning allows us to formulate the following statement, which provides new analytical bounds with controlled errors:

\begin{prop}

For arbitrary two Hermitian operators $X$ and $Y$, let $\Lambda(X)$ and $\Lambda(Y)$ denote the set of eigenvalues of the respective operators. Then, for the increasing finite sequences $(x)=(x_1,\ldots,x_n)$, $(y)=(y_1,\ldots,y_m)$ such that $\Lambda(X) \subset (x)$ and $\Lambda(Y) \subset (y)$, the following holds:
\begin{equation}
\Delta^2 X+\Delta^2 Y \ge \min_{i,j} \lambda_{\text{min}} (X_i+Y_j),
\end{equation}
where 
\begin{equation}
\eqalign{
X_i&=X^2-(x_i+x_{i+1})X+\mathbb{1}x_i x_{i+1} ,\\
Y_j&=Y^2-(y_j+y_{j+1})Y+\mathbb{1}y_j y_{j+1}.}
\end{equation}
The maximal error of approximation is bounded:
\begin{equation}
\Delta^2 X+\Delta^2 Y - \min_{i,j} \lambda_{\text{min}} (X_i+Y_j)\le \delta_X + \delta_Y,
\end{equation}
where
\begin{equation}
\delta_X=\left(\frac{\max_i (x_{i+1}-x_i)}{2}\right)^2~~~~\text{and}~~~~ 
\delta_Y=\left(\frac{\max_i (y_{i+1}-y_i)}{2}\right)^2.
\end{equation}
\label{prop:only}
\end{prop}

{This proposition and general idea about approximating uncertainty ranges generalizes naturally to higher number of observables: union of numerical ranges of sector decompositions is still the main object of interest.}

\section{Concluding remarks}
In this work we advocated a geometric approach to uncertainty relations, obtained state-independent tight analytical bounds applicable
 in low dimensions and semianalytical approximations for which maximal error can be controlled.
On one hand we presented a direct link between 
the algebraic notion
of numerical range of an operator \cite{toeplitz1918algebraische,hausdorff1919wertvorrat,horn1990matrix}
and uncertainty relations for the sum of variances \cite{maccone2014stronger}.
Furthermore, as numerical range of a matrix of order $d$ 
can be interpreted as a projection o the set  ${\cal M}_d$ of mixed quantum states 
of size $d$ on a plane, 
we have shown that uncertainty relations can be considered as 
a direct consequence of the highly non-trivial geometry of the set of quantum states. 

In particular, we applied the techniques described above to obtain exact analytical bounds for the sum of variances \eqref{eq:firstsum} of angular momentum operators for arbitrary total angular momentum quantum number $j$.
Furthermore, in Proposition \ref{prop:only} we provide an efficient method to generate arbitrarily precise approximations to uncertainty relations, applicable if exact calculations are too computationally expensive.

Our approach relies on geometric properties of the uncertainty region, the set of simultaneously attainable variances, which is generically not convex.
The method can be applied in entanglement detection schemes, analysis of security of quantum cryptographic 
protocols and diagnostics of quantum states.
Further generalization to state-dependent uncertainty relations is 
also possible with only minor modifications: 
along with calculation of the uncertainty range, averages of the observables of interest need to be determined.

Financial support by Narodowe Centrum Nauki under the project number
DEC-2015/18/A/ST2/00274 and Ministry  of  Science  and  Higher  Education  Grant  No. 0273/DIA/2016/4 is gratefully acknowledged. We would like to thank Alberto Riccardi for helpful remarks and suggestions.

\bigskip

\appendix
\section{Appendix A: Complementary approach using dual sets}

To proceed with an analytical approach it is convenient to work in the dual
space. 
This alternative description of the joint numerical range $W(X)$ is given by its \emph{dual set}. \emph{Dual} of an convex set $X$ can be defined as a set of linear functions $v$, for which the maximal over $X$ is equal to $1$, that is to say, $\sup_{\vec x\in X} v(\vec x)=1$. All these function may be viewed as calculating scalar products with some vectors, therefore we need to find vectors $\vec v$ such that $\sup_{\vec x\in X} \vec v\cdot \vec x=1$. This is the set $X^\circ$, dual to $X$:
\begin{equation}
\label{eq:dualdef}
X^{\circ}=\left\{ {\vec{v} } : \sup_{\vec x\in X} \vec{v} \cdot \vec x = 1 \right\} .
\end{equation}

Dual sets of numerical ranges admit an especially simple description as a set of roots of appropriately parametrized characteristic polynomial. This is a consequence of the relation between boundary of the numerical range $W(F_1,\ldots,F_n)$ and eigenvectors of mixed operators $\sum a_i F_i$.
\begin{figure}[H]
\centering
\includegraphics[width=0.5\columnwidth]{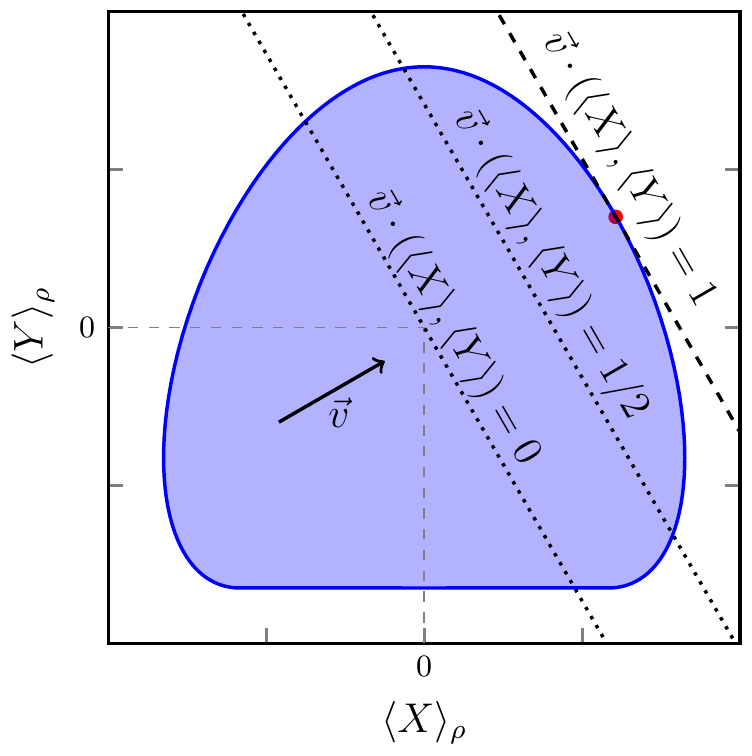}~~~~~%
\includegraphics[width=0.5\columnwidth]{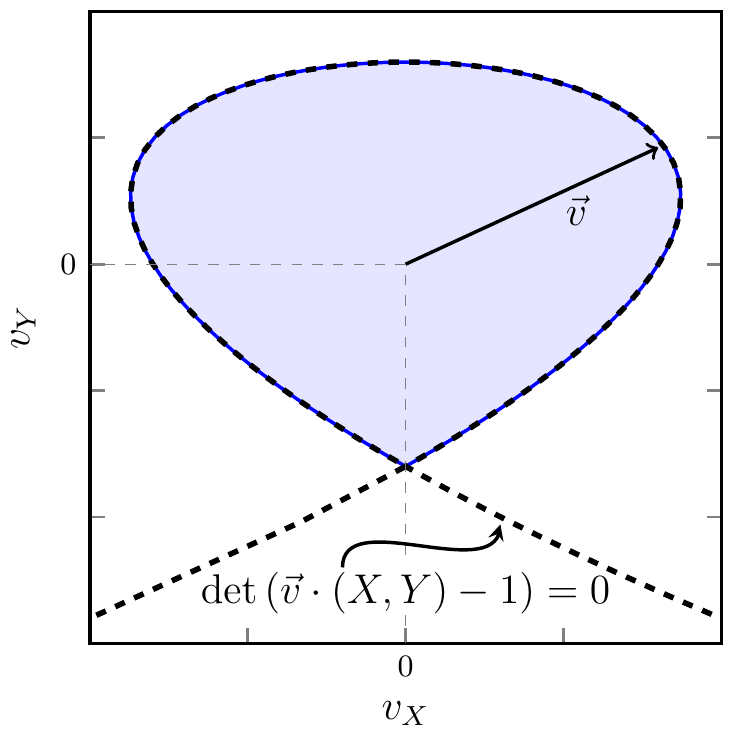}

\caption{Visualization of the process of finding the set $W^\circ$ (right panel), dual of the numerical range $W$ (left panel). The maximum of a scalar product is attained on an image of an eigenvector to the maximal eigenvalue, as shown on left panel.}

\label{fig:dual_example}
\end{figure}
As an example let us consider the numerical range of two operators $W(X,Y)$ -- we wish to calculate its dual set using the definition in Eq. \eqref{eq:dualdef}. One may ask: which vectors $\vec v$ proportional to $(\cos \theta, \sin \theta)$ belong to $W^\circ$? The defining criterion becomes then
\begin{equation}
\sup_{\rho} |v| (\cos \theta,\sin\theta)\cdot (\langle X\rangle, \langle Y\rangle)_\rho = 1,
\end{equation}
where the supremum is taken over all density matrices $\rho$. This is of course the same as asking that

\begin{equation}
\sup_{\rho} \langle X\cos \theta +Y\sin\theta \rangle_\rho = \frac1{|v|},
\end{equation}
that is to say, maximum eigenvalue of $X\cos \theta +Y\sin\theta$ is ${|v|}^{-1}$. Therefore, for every $\theta \in [0,2\pi)$, the point $(\cos\theta,\sin\theta) (\max\lambda(X\cos \theta +Y\sin\theta))^{-1}$ belongs to $W^\circ$. This implies the following constraint
\begin{equation}
\label{eq:vdet}
\vec v \in W^\circ \rightarrow \det \left( X v_x + Y v_y -1 \right) = 0.
\end{equation}
The reverse implication is \emph{not} true: $|v|$ may be inverse of any eigenvalue, including these which do not contribute to the boundary at all. There exists however unique way of determining which part of the curve defined by $\det \left( X v_x + Y v_y -1 \right) = 0$ belongs to $W^\circ$.

%

If the origin (zero vector) lies in the interior of $W(F_{1},\ldots,F_{n})$,
the subset of the affine variety forming $W^{\circ}$ can be easily
identified -- since the boundary of joint numerical range is defined by \emph{maximal} eigenvalues, if we choose any direction $\vec n\in S^{k-1}$ and take all solutions of the right side in \eqref{eq:vdet} of the form $\vec v=\lambda \vec n$ ($\lambda>0$), the boundary $\partial W$ of JNR is described by the one with \emph{minimal} $\lambda$. The set $W^\circ$ is therefore a kind of a `cell' around the origin of the dual space.

In previous sections the function $g=\left\langle X^2+Y^2\right\rangle 
 -\left\langle X\right\rangle ^{2}-\left\langle Y\right\rangle ^{2}$ and a method to generate the set of all triples of expectation values $(\langle X\rangle, \langle Y\rangle, \langle X^2+Y^2\rangle)$ have been introduced. Here we demonstrate how to leverage the algebraic structure of numerical range $W(X,Y,X^2+Y^2)$ in the dual space to determine the tight, analytical bounds for the sum of variances (Eq. \eqref{eq:firstsum}). 
 
As a starting point we can assume that without loss of generality $\Tr X=\Tr Y=0$ --- constant shift does not change the variances. Furthermore,
let us introduce a traceless operator $Z$ related to the sum of squares, 
\begin{equation}
Z=X^{2}+Y^{2}-\frac{\Tr (X^{2}+Y^{2})}{d}\mathbb{1},
\end{equation}
where $d$ denotes the dimension of the Hilbert space. 
Minimizing the function $g'=\left\langle Z\right\rangle -\left\langle X\right\rangle ^{2}-\left\langle Y\right\rangle ^{2} + \Tr(X^2+Y^2)/d$ is equivalent to the original problem:
 the result $C_{\text{min}}$ yields the desired bound for sum of variances. Working with traceless operators later proves to be convenient.
 
The problem can also be rephrased in terms of geometry:
For a given convex set $W(X,Y,Z)\subset \mathbb{R}^3$ 
calculate the minimal shift $t$ of the paraboloid of a constant uncertainty, $P_\tau=\{(x,y,z)\in\mathbb{R}^3 : z-x^{2}-y^{2}=\tau\}$, 
such that the paraboloid $P_\tau$ is tangent to the numerical range $W$
-- see Fig. \ref{fig:uncermac}.

\begin{figure}[t]
\centering
\includegraphics[width=0.6\columnwidth]{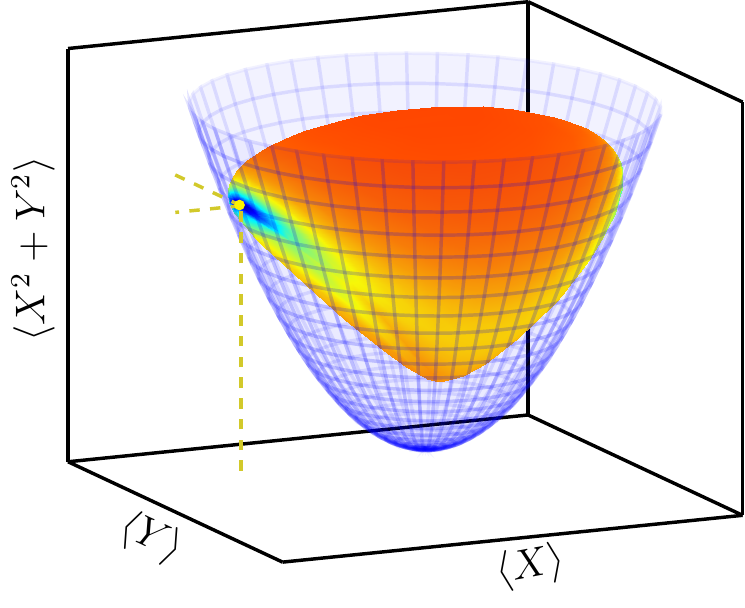}
\caption{3D numerical range $W(X,Y,X^2+Y^2)$ for $X$ and $Y$ defined in Eq. \eqref{eq:nonmom}, shaded according to sum of variances $\Delta^2 X+\Delta^2 Y$.
Paraboloid of a minimal uncertainty (sum of variances) is shown, as well as the minimizing point (yellow dot and dashed guiding lines). }
\label{fig:uncermac2}
\end{figure}

The theory of numerical range implies \cite{henrion2010semidefinite} 
that the dual set $W^\circ$ is contained in the set of roots of 
the polynomial in real variables $u$, $v$ and $w$:
\begin{equation}
Q(u,v,w):=\det(uX+vY+wZ-\mathbb{1})=0.
\end{equation}

Tracelessness of all operators come in handy during the analysis of this set: 
we exactly know what subset of solutions \kon{of} this equation \kon{forms} $W^\circ$. From the definition involving dual space it is also apparent why to minimalize Eq. \eqref{eq:form4} we can assume smooth behavior of the function, as in Eq.\eqref{eq:f2d}: the cusps in dual the space correspond to flat surfaces in real space: variance, being a strictly convex function, does not attain minimum on a flat part of numerical range.

The paraboloid is not a numerical range, however, its dual can be defined analogously by  relation \eqref{eq:dualdef}. It admits the same analytical treatment aimed at determination of the defining polynomial in dual space: the dual of the paraboloid $P_\tau$ for a fixed parameter $\tau$, 
denoted by $P_\tau^\circ$, is determined by equation
\begin{equation}
D(\tau):=u^{2}+v^{2}+4w-4w^{2}\tau=0.
\label{eqn:paradual}
\end{equation}

For a negative $\tau$, 
the set of roots of \eqref{eqn:paradual} becomes a hollow ellipsoid. We are going to seek the roots in this region, since the operator $Z$ is $X^2+Y^2$ shifted by identity matrix by $-\Tr(X^2+Y^2)/d$; let us denote this value by $\lambda_0$ for further usage.

The bound $C$ for the sum of variances, determined by the condition
that both manifolds intersect,
is thus obtained by solving the 
following set of polynomial equations for real variables 
$u,v,w,\lambda$ 
and the multiplier $\alpha$,
\begin{equation}
\left\{\begin{array}{rl}
D(\lambda-\lambda_0)& =0,\\
Q(u,v,w)& =0, \\
\partial_i (Q(u,v,w)-\alpha D(\lambda-\lambda_0)) &=0~~\text{for}~i=u,v,w.
\end{array}\right.
\label{eq:ddd}
\end{equation} 
It is always possible to find a polynomial $R(\lambda)$, the roots of which 
correspond to local extrema of variance -- see Appendix B.
The bound corresponds to the minimal \emph{real} root of the polynomial equation
$R(\lambda)=0$, denoted by $\lambda_{\text{min}}$,
equivalent to the above set of five equations, 
for which real solutions to $u,v,w,\alpha$ exist (in general they may be complex regardless of real $\lambda$).
Then the tight bound for $\Delta^2 X+\Delta^2 Y$ reads
\begin{equation}
\Delta^{2}X+\Delta^{2}Y\ge  C=\lambda_{\text{min}}.
\label{eq:ddd2}
\end{equation}
It is possible to determine the state saturating the uncertainty relation by solving the polynomial system described in \eqref{eq:ddd} with variable $\lambda$ set to the calculated constant $\lambda_{\text{min}}$.

A general solution of the problem in the simplest 
case $d=2$  is provided in Appendix C, while for higher dimensions,
calculations performed for individual cases 
yield semianalytical results determined by roots of a particular polynomial.

\section{Appendix B: Solving the equations leading to uncertainty relation}
The simplification of the set of equations {\eqref{eq:ddd}} or \eqref{eq:f2d} may be found in an algebraic way, which presents the result in an explicitly analytical way. This is the method of Gr\"obner basis, described below. 

Consider arbitrary set of polynomial equations of finite number of variables:

\begin{equation}
\left\{\begin{array}{rl}
P_1(x_1,\ldots,x_k)& =0,\\
P_2(x_1,\ldots,x_k)& =0,\\
\ldots&\\
P_n(x_1,\ldots,x_k)& =0.\\
\end{array}\right.
\label{eq:nwm}
\end{equation}

Our goal is to calculate the realization of variables $(x_1,\ldots,x_k)$ with minimal $x_1$ among all real solutions to this system of equations. Let us assume that the set of \emph{all} solutions -- which may be complex -- is discrete. It is the generic case if number of equations $n$ is equal to number of variables $k$ -- a result reminiscent to the linear algebra in which a matrix equation $A \vec x=\vec y$ has a single solution if matrix $A$ is nondegenerate.

The algorithm used to determine set of common solutions to \eqref{eq:nwm} has some resemblance to Gaussian elimination known from linear algebra. To solve the system of linear equations we
\begin{enumerate}
\item choose an order of variables, i.e. any permutation of $(x_1,\ldots,x_k)$ interpreted as sequence of symbols,
\item recursively reduce system of equations by mutliplication and addition, such that at the end a \emph{single equation} for the \emph{last variable} in the chosen order is obtained (e.g. for a natural order $(x_1,\ldots,x_k)$ we get a linear equation for $x_k$),
\item recursively solve the equations for remaining variables by back substitution.
\end{enumerate}

Conceptually, the \emph{Buchberger algorithm} \cite{cox1992ideals} which solves the system of polynomial equations \eqref{eq:nwm} does not differ much:

\begin{enumerate}
\item a choice of order of \emph{monomials} -- expressions of form $x_1^{\alpha_1}x_2^{\alpha_2}\ldots x_k^{\alpha_k}$ -- is needed. This order will determine the last remaining equation to solve, just as in the case of variables in Gaussian elimination. Since exponents $\alpha_1,\ldots,\alpha_k$ in monomials may potentially become very large, a consistent and simple to calculate order choice is needed. A natural choice is such that the last elements of order are powers of $x_k$ only.
\item The system of polynomial equations is recursively reduced by mutliplication by polynomials and addition, such that in each step a \emph{leading term} (defined by chosen order) of polynomials involved is decreasing. If we choose the order described above, the  \emph{last polynomial equation} consists of monomials in $x_k$ only.
\item The last equation -- a polynomial in $x_k$ only -- is solved; remaining equations are recursively solved by back substitution.
\end{enumerate}

In our case we wish to set $\lambda$ of appearing in Eq. \eqref{eq:ddd} or Eq. \eqref{eq:f2d} as the last variable, so that all others ($x,y$ in Eq. \eqref{eq:ddd}, $u,v,w,\alpha$ in Eq. \eqref{eq:f2d}) do not appear in the final solution. The minimal real root of the resulting polynomial in $\lambda$ for which all the other corresponding variables also have real values is exactly what we are looking for in the first place.


\section{Appendix C: Numerical range and bounds for $\bf d=2$}
In the case of qubit observable $X$, the squared operator $X^2$ may always be written as combination of identity and $X$: $X^2=\frac{\mathbb{1}}{2} \Tr X^2  +\frac12 X \Tr X$. Additionally, the joint numerical range of two operators of size $d=2$ is an ellipse, hence the joint numerical range of three operators $W(X,Y,X^2+Y^2)$ is a flat object -- linear transformation of ellipse. The variance is thus minimized on the curve bounding the joint numerical range. This is already known \cite{abbott2016tight}; here we reconstruct the result using the analytical method of solving the Eqns. \eqref{eq:ddd}. 

Arbitrary qubit observables $X$ and $Y$ can always be rescaled to form
\begin{equation}
X=\left(\begin{array}{cc}1&0\\0&-1\end{array}\right),~~Y=\left(\begin{array}{cc}a&b\\b^*&-a\end{array}\right).
\end{equation}
The resulting polynomial reads
\begin{equation}
a^2 t+|b|^2 t+|b|^2+t^2+t,
\end{equation}
and implies bound $C$ for the sum of variances
\begin{equation}
C=\frac{1}{2} \left(a^2+\left| b\right| ^2+1 - \sqrt{\left(a^2+\left| b\right| ^2+1\right)^2-4 |b|^2}\right).
\end{equation}

\section{Appendix D: Resulting polynomials}

In Table \ref{tab:polys} we present some of the resulting polynomials defining the minimal sums of variances presented in Table 1. The polynomials for higher total angular momenta ($j>4$) are omitted due to their length. The only discernible pattern we have observed is in the order of polynomials -- the online catalogue OEIS recognizes this sequence as part of A243099, with general term (for $n\ge 1$)
\begin{equation}
o(n)=\frac{1}{16} \Big(6+n\big(3 n+2\big)-(-1)^{n} \big(n\left(n-2\right)+6\big)\Big).
\end{equation}
\begin{table}[H]
\begin{tabular}{c|l}
Total angular \\momentum $j$& polynomial $P(\lambda)$ for which $P(\min \Delta^2 J_X+\Delta^2 J_Y)=0$\\\hline
$\frac12$& $4\lambda-1$\\\hline
1&$16\lambda-7$\\\hline
$\frac32$& $64 \lambda^3-336 \lambda^2+480 \lambda -181$\\
\hline
2&$1024 \lambda ^3-7104 \lambda ^2+13404 \lambda -6487$\\\hline
$\frac52$& $\begin{array}{l}
4194304 \lambda^7-117440512 \lambda^6+1323466752 \lambda^5\\
-7743660032 \lambda^4\\
+25301870144 \lambda^3-45976348848 \lambda^2+42609045676 \lambda -15158613241\end{array}$\\
\hline
3&$%
\begin{array}{l}
262144 \lambda ^6-8159232 \lambda ^5+98042880 \lambda ^4-574842880 \lambda ^3+1709341632 \lambda ^2\\
-2397898539 \lambda +1179352998
\end{array}$\\\hline
$\frac72$ & $\begin{array}{l}8589934592 \lambda ^{13}-807453851648 \lambda ^{12}+33978291585024 \
\lambda ^{11}-846111845646336 \lambda^{10}\\+13886438674268160 \lambda^9-158338610153127936 \lambda^8+1288197712964943872 \lambda^7\\-7558113438967267328 \lambda^6+31923050400995246592 \lambda^5\\
-95672589723220763904 \lambda^4+197231954550318498240 \lambda^3\\
-263781099083569171968 \lambda^2+203719751641654010210 \lambda
\\-67745907126251250695\end{array}$
\\\hline
4&$\begin{array}{l}268435456 \lambda^{10}-24326963200 \lambda^9+952177262592 \lambda \
^8-21110360768512 \lambda^7\\
+292217512296448 \lambda^6-2624974965550080 \lambda^5+15404738985045728 \lambda^4\\
-57914832267046937 \lambda^3+132277607024648928 \lambda^2\\
-163205035294297553 \lambda +79926677043771116\end{array}$
\end{tabular}
\caption{Table of polynomials defining strict minimal values of sum of variances for angular momentum operators. The polynomials are \emph{minimal}: there exist no lower order polynomials with integer coefficients for which $\min \Delta^2 J_X+\Delta^2 J_Y$ is one of the roots.}

\label{tab:polys}
\end{table}


\bigskip 

\bibliographystyle{unsrtabbrv}
\bibliography{bibliography}

\begin{thebibliography}{10}

\bibitem{scully1991quantum}
Scully, M.~O., Englert, B.-G., and Walther, H.
\newblock {\em Nature}, 351(6322):111--116, 1991.

\bibitem{tomamichel2012tight}
Tomamichel, M., Lim, C. C.~W., Gisin, N., and Renner, R.
\newblock {\em Nat. Commun.}, 3:634, 2012.

\bibitem{hofmann2003violation}
Hofmann, H.~F. and Takeuchi, S.
\newblock {\em Phys. Rev. A}, 68(3):032103, 2003.

\bibitem{guhne2004characterizing}
G{\"u}hne, O.
\newblock {\em Phys. Rev. Lett.}, 92(11):117903, 2004.

\bibitem{sorensen2001entanglement}
S{\o}rensen, A.~S. and M{\o}lmer, K.
\newblock {\em Phys. Rev. Lett.}, 86(20):4431, 2001.

\bibitem{kennard1928note}
Kennard, E.
\newblock {\em Phys. Rev.}, 31(3):344, 1928.

\bibitem{heisenberg1930physikalische}
Heisenberg, W.
\newblock Physikalische prinzipien der quantentheorie (Leipzig: Hirzel), 1930.

\bibitem{maccone2014stronger}
Maccone, L. and Pati, A.~K.
\newblock {\em Phys. Rev. Lett.}, 113(26):260401, 2014.

\bibitem{bialynicki1984entropic}
Bialynicki-Birula, I.
\newblock {\em Phys. Rev. A}, 103(5):253--254, 1984.

\bibitem{schwonnek2017state}
Schwonnek, R., Dammeier, L., and Werner, R.~F.
\newblock {\em Phys. Rev. Lett.}, 119(17):170404, 2017.

\bibitem{giorda2018state}
Giorda, P., Maccone, L., and Riccardi, A.
\newblock {\em arXiv preprint arXiv:1810.09775}, 2018.

\bibitem{sehrawat2017deriving}
Sehrawat, A.
\newblock {\em arXiv preprint arXiv:1706.09319}, 2017.

\bibitem{dammeier2015uncertainty}
Dammeier, L., Schwonnek, R., and Werner, R.~F.
\newblock {\em New Journal of Physics}, 17(9):093046, 2015.

\bibitem{szymanski2017uncertainty}
Szyma{\'n}ski, K.
\newblock {\em arXiv preprint arXiv:1707.03464}, 2017.

\bibitem{toeplitz1918algebraische}
Toeplitz, O.
\newblock {\em Math. Z.}, 2(1):187--197, 1918.

\bibitem{hausdorff1919wertvorrat}
Hausdorff, F.
\newblock {\em Math. Z.}, 3(1):314--316, 1919.

\bibitem{horn1990matrix}
Horn, R.~A. and Johnson, C.~R.
\newblock {\em Matrix analysis}.
\newblock Cambridge {U}niversity {P}ress, 1990.

\bibitem{keeler1997numerical}
Keeler, D.~S., Rodman, L., and Spitkovsky, I.~M.
\newblock {\em Linear Algebra Its Appl.}, 252(1-3):115--139, 1997.

\bibitem{helton2011possible}
Helton, J.~W. and Spitkovsky, I.~M.
\newblock {\em arXiv preprint arXiv:1104.4587}, 2011.

\bibitem{li2000convexity}
Li, C.-K. and Poon, Y.-T.
\newblock {\em SIAM Journal on Matrix Analysis and Applications},
  21(2):668--678, 2000.

\bibitem{szymanski2016classification}
Szyma{\'n}ski, K., Weis, S., and {\.Z}yczkowski, K.
\newblock {\em Linear Algebra Its Appl.}, 545:148--173, 2018.

\bibitem{chen2016physical}
Chen, J.-Y., Ji, Z., Liu, Z.-X., Qi, X., Yu, N., Zeng, B., and Zhou, D.
\newblock {\em SCIENCE CHINA Physics, Mechanics \& Astronomy}, 60(2):020311,
  2017.

\bibitem{zauner2016njp}
Zauner-Stauber, V., Draxler, D., Vanderstraeten, L., Haegeman, J., and
  Verstraete, F.
\newblock {\em New J. Phys.}, 18(11):113033, 2016.

\bibitem{spitkovsky2017new}
Spitkovsky, I.~M. and Weis, S.
\newblock {\em arXiv preprint arXiv:1703.00201}, 2017.

\bibitem{oeis}
{OEIS Foundation Inc.}
\newblock The On-Line Encyclopedia of Integer Sequences.
\newblock {http://oeis.org}, 2019.

\bibitem{hadwiger1950minkowskische}
Hadwiger, H.
\newblock {\em Math. Z.}, 53(3):210--218, 1950.

\bibitem{henrion2010semidefinite}
Henrion, D.
\newblock {\em Electron. J. Lin. Algebra.}, 2:1--2, 2010.

\bibitem{cox1992ideals}
Cox, D., Little, J., and O'{S}hea, D.
\newblock {\em Ideals, {V}arieties, and {A}lgorithms}, volume~3.
\newblock Springer, 1992.

\bibitem{abbott2016tight}
Abbott, A.~A., Alzieu, P.-L., Hall, M.~J., and Branciard, C.
\newblock {\em Mathematics}, 4(1):8, 2016.

\end{thebibliography}

\end{document}